\documentclass{PoS}
\newcommand{\ber}{\begin{eqnarray}}
\newcommand{\eer}[1]{\label{#1}\end{eqnarray}}
\newcommand{\re}[1]{(\ref{#1})}
\newcommand{\bbD}[1]{\mathbb{D}_{#1}}
\newcommand{\bbDB}[1]{\bar{\mathbb{D}}_{#1}}
\newcommand{\bbX}[1]{\mathbb{X}^{#1}}

\def\+{{+\!\!\!+}}
\def\pp{\mbox{\tiny${}_{\stackrel\+ =}$}}

\title{Complex Geometry and Supersymmetry}

\ShortTitle{Complex Geometry and Susy}

\author{\speaker{Ulf Lindstr\"om}\\
        Division of Theoretical Physics\\
        Department of Physics and Astronomy\\
        Uppsala University\\
        SWEDEN\\
        E-mail: \email{ulf.lindstrom@physics.uu.se}}

\abstract{I stress how the form of sigma models with $(2,2)$ supersymmetry differs depending on the number of  manifest supersymmetries.\
          The differences correspond to different aspects/formulations of Generalized K\"ahler Geometry.}

\FullConference{Proceedings of the Corfu Summer Institute 2011 School and Workshops on Elementary Particle Physics and Gravity\\
		September 4-18, 2011\\
		Corfu, Greece}

\begin{document}

\section{Introduction}
In this brief presentation I report on an aspects of the relation between twodimensional $N=(2,2)$ sigma models and complex geometry that I find remarkable:
To each superspace formulation of the sigma model, be it  $N=(2,2)$, $N=(2,1)$, $N=(1,2)$ or $N=(1,1)$, there is always  a natural corresponding formulation of the Generalized K\"ahler Geometry  on the target space. I first introduce the relevant formulations of Generalized K\"ahler Geometry and then the sigma models. The results are collected from a number of papers where we have used sigma models as tools to probe the geometry: \cite{Lindstrom:2004eh}-\cite{Hull:2012dy}. See also \cite{Sevrin:1996jr}, \cite{Bogaerts:1999jc} for  related early discussions.

\section{Formulations of Generalized K\"ahler Geometry}
Generalized K\"ahler Geometry was defined by Gualtieri \cite{Gualtieri:2003dx} in his PhD thesis on Generalized Complex Geometry. The latter subject was introduced by Hitchin in \cite{Hitchin:2004ut}. In \cite{Gualtieri:2003dx} it is also described how GKG is a reformulation of the bihermitean geometry of \cite{Gates:1984nk}, which we now turn to.
\subsection{Generalized K\"ahler Geometry I; Bihermitean Geometry.}
Bihermitean geometry is the set $(M, g, J_{(\pm)}, H)$, i.e., a manifold $M$ equipped with a metric $g$, two complex structures $J_{(\pm)}$ and a closed three-form $H$. The defining properties may be summarized as follows:
  \bigskip

\begin{table}[htdp]
\begin{center}
\begin{tabular}{|c|}
\hline
\cr
$J_{(\pm)}^2=-1\!\!\!1~, \quad J_{(\pm)}^tgJ_{(\pm)}=g~,\quad \nabla^{(\pm)}J_{(\pm)}=0$\cr
\cr
 $\Gamma^{(\pm)}=\Gamma^0\pm\frac 1 2  g^{-1}H~,\quad H=dB~.$\cr
 \cr
\hline
\end{tabular}
\end{center}
\label{bih1}
\caption{Bihermitean 1}
\end{table}
\noindent
In words, the metric is hermitean with respect to \underline{both} complex structures and these, in turn, are covariantly constant with respect to  connections which are the sum of the Levi-Civita connection and a torsion formed from the closed three-form. Locally, the three-form may be expressed in terms of a potential two-form $B$. This $B$-field, or NSNS two-form, is conveniently combined with the metric into one tensor $E$:
\ber
E:=g+B~.
\eer{0}
\bigskip

A reformulation of the data in Table.1 more adapted to Generalized Complex Geometry is as the set {$(M, g, J_{(\pm)})$} supplemented with (integrability)conditions acoording to
\eject

\begin{table}[htdp]
\begin{center}
\begin{tabular}{|c|}
\hline
\cr
$J_{(\pm)}^2=-1\!\!\!1~, \quad J_{(\pm)}^tgJ_{(\pm)}=g~,\quad \omega_{(\pm)}:=gJ_{(\pm)}$\cr
\cr
 $d^c_{(+)}\omega_{(+)}+d^c_{(-)}\omega_{(-)}=0~,\quad dd^c_{(\pm)}\omega_{(\pm)}=0~,$\cr
 \cr
 $ H:=d^c_{(+)}\omega_{(+)}=-d^c_{(-)}\omega_{(-)}$\cr
 \cr
\hline
\end{tabular}
\end{center}
\label{bih2}
\caption{Bihermitean 2}
\end{table}
\noindent
Here $\omega_\pm$ are the generalizations of the K\"ahler forms for the two complex structures, $d^c$ is the differential which reads $i(\bar \partial-\partial)$ in local coordinates where the complex structure is diagonal, and we see that the three-form is defined in terms of the basic data.\bigskip

\subsection{Generalized K\"ahler Geometry II; Description on $T\oplus T^*$}

Generalized Complex Geometry  \cite{Hitchin:2004ut}, \cite{Gualtieri:2003dx}, is formulated on the sum of the tangent and cotangent bundles $T\oplus T^*$ equipped with an endomorphism which is a (generalized) almost complex structure, i.e., a map 
\ber
{\cal J}:T\oplus T^*\to T\oplus T^*~:~~
{\cal J}^2=-1\!\!\!1~.
\eer{}
The further requirements that turn ${\cal J}$ into a generalized complex structure is first that it preserves the natural pairing on $T\oplus T^*$
\ber
{\cal J}^t{ I}{\cal J}={ I}~, ~~~~I:=\left(\begin{array}{ccc}0&1\cr1&0\end {array}\right)~,
\eer{}
where the matrix expression refers to the coordinate basis $(\partial_\mu, dx^\nu)$ in $T\oplus T^*$, and second the integrability condition
\ber
\pi_\mp\left[\pi_\pm X, \pi_\pm Y\right]_C = 0~, ~~~~X,Y \in T\oplus T^*.
\eer{cb}
Here $C$ denotes the Courant bracket \cite{Courant}, which for $X=x+\xi,Y=y+\eta \in T\oplus T^*$ reads
\ber
\left[ X,  Y\right]_C:=[x,y]+{\cal L}_x\eta-{\cal L}_y \xi-\frac 1 2 d(\imath_x\eta-\imath_y\xi)~,
\eer{}
with the Lie bracket, Lie derivative and contraction of forms with vectorfields appearing on the right hand side. Generalized K\"ahler Geometry \cite{Gualtieri:2003dx} requires the existence of \underline{two commuting} such Generalized Complex Structures, i.e.:
\ber{\cal J}_{(1,2)}^2=-1\!\!\!1\quad [{\cal J}_{(1)},{\cal J}_{(2)}]=0~,\quad {\cal J}_{(1,2)}^t{ I}{\cal J}_{(1,2)}={ I}~, \quad {\cal G}:=-{\cal J}_{(1)}{\cal J}_{(2)}~,
\eer{1}
with both GCSs satisfying \re{cb} and the last line defines an almost product structure ${\cal G}$:
\ber
{\cal G}^2=1\!\!\!1~.
\eer{}
When formulated in $T\oplus T^*$, K\"ahler geometry satisfies these condition, and so does bihermitean geometry. In fact the Gualtieri map \cite{Gualtieri:2003dx} gives the precise relation\footnote{The derivation from sigma models is given in \cite{Bredthauer:2006hf}.} to the data in Table 2: 

\ber{\cal J}_{(1,2)}=\left(\begin{array}{ccc}1&&0\cr &&\cr B&&1\end {array}\right)\left(\begin{array}{ccc}J_{(+)}\pm J_{(-)}&&-(\omega_{(+)}^{-1}\mp\omega_{(-)}^{-1})\cr&&\cr\omega_{(+)}\mp\omega_{(-)}&&-(J_{(+)}^t\pm J_{(-)}^t)\end {array}\right)\left(\begin{array}{ccc}1&&0\cr &&\cr-B&&1\end {array}\right)
\eer{2}
\bigskip

\subsection{Generalized K\"ahler Geometry III;  Local Symplectic Description }
Bihermitean geometry emphasizes the complex aspect of  generalized K\"ahler Geometry. There is another formulation where the (local) symplectic structure is in focus.

Given the bi-complex manifold {$(M, J_{(\pm)})$}, there  exists
locally defined non-degenerate ``symplectic'' two-forms ${\cal F}_{(\pm)}$ such that $d{\cal F}_{(\pm)}=0$ and \cite{Hull:2008vw}
\begin{table}[htdp]
\begin{center}
\begin{tabular}{|c|}
\hline
\cr
${\cal F}_{(\pm)}(v,J_{(\pm)}v)>0~, \quad d({\cal F}_{(+)}J_{(+)}-J_{(-)}^t{\cal F}_{(-)})=0~.$\cr
\cr
\hline
\end{tabular}
\end{center}
\label{default}
\caption{Conditions on ${\cal F}$}
\end{table}

In the first condition $\nu$ is an arbitrary contravariant vector field and the condition says that  ${\cal F}_\pm$ tames the complex structures $J_{(\pm)}$. The bihermitean data is recovered from
\ber\nonumber
&&{\cal F}_{(\pm)}=\frac 12i (B^{(2,0)}_{(\pm)}-B^{(0,2)}_{(\pm)})\mp\omega_{(\pm)}\\[1mm]
&&{\cal F}_{(+)}=-\frac 12 E^t_{(+)}J_{(+)}~,\quad {\cal F}_{(-)}=-\frac 12 J_{(-)}^tE_{(-)}^t
\eer{3}
where, e.g.,  $B_{(\pm)}^{(2,0)}$ refers to the holomorphic property of $B$ under $J_{(\pm)}$.

 \subsection{Summary}
 \bigskip

As we have seen, the geometric data representing Generalized K\"ahler Geometry may be packaged in various equivalent ways as, e.g.,
 $(M,g,H,J_{(\pm)})$, as  $(M,g,J_{(\pm)})$ or as $(M,{\cal F}_{(\pm)},J_{(\pm)})$.
In each case, there is a complete description in terms of a Generalized K\"ahler potential $K$  \cite{Lindstrom:2005zr}\footnote{The description is complete away from irregular points of certain poisson structures}.
Unlike the K\"ahler case, the expressions are non-linear in second derivatives of $K$. E.g., restricting attention to the situation $[J_{(+)},J_{(-)}]\ne \emptyset$, the left complex structure is given by
\ber
J_{(+)}=\left(\begin{array}{ccc}J&&0\cr &&\cr (K_{LR})^{-1}[J,K_{LL}]&&(K_{LR})^{-1}JK_{LR}\end {array}\right)~,
\eer{4}
where we we introduced local coordinates $(\bbX{L},\bbX{R})~,~~L:= \ell,\bar\ell~,~R:=r,\bar r$, and $K_{LR}$ is shorthand for the matrix
\ber
K_{LR}:=\left( \begin{array}{ccc}
\frac{\partial^2 K}{\partial\bbX{\ell}\partial\bbX{r}}&&\frac{\partial^2 K}{\partial\bbX{\ell}\partial\bbX{\bar r}}\\
&&\\
\frac{\partial^2 K}{\partial\bbX{\bar\ell}\partial\bbX{r}}&&\frac{\partial^2 K}{\partial\bbX{\bar\ell}\partial\bbX{\bar r}}\end{array}\right)~.
\eer{44}
The metric is
\ber
g=\Omega[J_{(+)},J_{(-)}]~,
\eer{5}
and the local symplectic structures have potential one-forms $\lambda_{(\pm)}$. E.g.,
\ber
{\cal F}_{(+)}=d\lambda_{(+)}~, \quad \lambda_{(+)\ell} = iK_RJ(K_{LR})^{-1}K_{L\ell}~,...
\eer{6}

The  relations may be extended to the whole manifold in terms of gerbes  \cite{Hull:2008vw}.

\section{Sigma Models}{}
The $d=2~,~N=(2,2)$ supersymmetry algebra of covariant derivatives is
\ber \{\bbD{\pm},\bbDB{\pm}\}=i\partial_{\pp}
\eer{}
The covariant derivatives can be used to constrain superfields. We shall need chiral, twisted chiral and left and right semichiral superfields \cite{Buscher:1985kb}:
\ber\nonumber
\bbDB{\pm}\phi^a=0~,\\[1mm]\nonumber
\bbDB{+}\chi^{a'}=\bbD{-}\chi^{a'}=0~,\\[1mm]\nonumber
\bbDB{+}\bbX{\ell}=0~,\\[1mm]
\bbDB{-}\bbX{r}=0~,
\eer{77}
and their complex conjugate.
The collective indexnotation is taken to be; $c:=a~,\bar a,~~ t:=a',\bar a',$ and, as before, $L:= \ell,\bar\ell,~~ R:=r,\bar r.$
\subsection{Superspace I}
The $(2,2)$ formulation of the $(2,2)$ sigma model uses the generalized K\"ahler Potential $K$ directly:
\ber
S=\int \bbD{+}\bbDB{+}\bbD{-}\bbDB{-}K(\phi^c,\chi^t,\bbX{L},\bbX{R})
\eer{7}
 Note that  $K$ has  many roles: as a Lagrangian as in \re{7}, as a potential for the geometry, \re{4}, \re{44}, as a ``prepotential'' for 
the local symplectic form ${\cal F}$, \re{6}, and, as shown in \cite{Lindstrom:2005zr}, as a generating function for symplectomorphisms 
between coordinates where 
$J_{(+)}$ and  coordinates where $J_{(-)}$ are canonical.

\subsection{Superspace II}
To discuss reduction of the action \re{7} to $(2,1)$ superspace \cite{Hull:2012dy},
we restrict the potential to $ K(\bbX{L},\bbX{R})$ to simplify the expressions.

The reduction entails representing the $(2,2)$ right derivative as a sum of $(2,1)$ derivative and a generator of supersymmetry:
\ber
\bbD{-}=:D_--iQ_-~, 
\eer{} 
and defining the $(2,1)$ components of a $(2,2)$ superfield as
\ber
\bbX{}|=:X~, ~~ Q_-\bbX{L}|=:\Psi_-^L~.
\eer{}
The action \re{7} then reduces as
\ber
S=\int \bbD{+}\bbDB{+}D_{-}\left(K_L\Psi_-^L+K_RJD_-X^R\right)~.
\eer{88}
Here $\Psi$ is a Lagrange multiplier field enforcing $\bbDB{+}K_\ell=0$ and its c.c., which are the $(2,1)$ components of the $(2,2)~~\bbX{\ell}$ and
 $\bbX{\bar\ell}$ equations.  We solve this by going to $(2,2)$ coordinates $(\bbX{L}, \mathbb{Y}_L)$ \cite{Lindstrom:2005zr}, \cite{Hull:2012dy}, whose $(2,1)$ components will now both be chiral. The action then reads
 \ber
&&S=i\int \bbD{+}\bbDB{+}D_{-}(\lambda_{(+)\alpha}D_-\varphi^\alpha+c.c.)
\eer{8}
with $\varphi^\alpha \in (X^\ell,Y_\ell)$ and $\bbDB{+}\varphi^\alpha=0$. This is the standard form of a $(2,1)$ sigma model \cite{AbouZeid:1997cw} but with the vector potential now identified (up to factors) with $\lambda_{(+)}$ in \re{6}, (${\cal F}_{(+)}=d\lambda_{(+)}$). Of the two complex structures 
$J_{(\pm)}$ only $J_{(+)}$ is now manifest. The complex structure $J_{(-)}$ instead appears in the non-manifest supersymmetry
\ber
\delta \varphi ^\alpha=\bbDB{+}(\epsilon J^{~\alpha}_{(-)i}D_-\phi^i)~,~~~\{\phi^i\}=\{\varphi^\alpha,\bar\varphi^{\bar\alpha}\}
\eer{41}

Similarily, reduction of \re{7} to $(1,2)$ yields a model in which $J_{(-)}$ is the remaining manifest complex structure. It is found from the $(2,1)$ model by the replacement $+\to -$, and  $L\to R$.

\subsection{Superspace III}

We may reduce the action \re{7} to $(1,1)$ superspace directly or via the $(2,1)$ formulation. The resulting action now involves the metric and $B$-fields in the combination \re{0} as geometric objects:
\ber
S=\int D_{+}D_{-}\left(D_+XED_-X\right)~,
\eer{9}
where we have supressed the indices.
Starting from $(2,1)$ superspace and the action \re{88}, the reduction goes via
\ber
\bbD{+}=:D_+-iQ_+~,\quad Q_+\bbX{R}|=:\Psi_+^R~, 
\eer{}
and  both the auxiliary spinors $\Psi_-^L$ and $\Psi_+^R$ have been eliminated\footnote{Note that these spinors have the role of Lagrange multipliers in the 
$(2,1)$ and $(1,2)$ formulations, but the role of auxiliary fields with algebraic field equations in the $(1,1)$ formulation}. Both complex structures are now non-manifest and arise in the extra supersymmetry transformations as explained in \cite{Gates:1984nk}.

\subsection{Summary}

The various sigma models have different formulations of Generalized K\"ahler Geometry manifest.
Thus the  $(2,2)$ sigma model is written directly in terms of the generalized K\"ahler potential. The  $(2,1)$ or  $(1,2)$ model involves the one form
$\lambda_{(+)}$ or $\lambda_{(-)}$ respectively, which connects it to the local symplectic formulation. The $(1,1)$ sigma model, finally, is expressed directly in terms of the metric and $B$-field, making that aspect of the geometry manifest. These are also the objets that determine the $(0,0)$ formulation, i.e., the component formulation of the sigma model.
\bigskip

\noindent
{\bf Acknowledgement}\\{This short presentation highlights some aspects of the relation between sigma models and Generalized K\"ahler Geometry derived in a long and ongoing collaboration with Chris Hull, Martin Ro\v cek,  Rikard von Unge, Maxim Zabzine and others. I am grateful to all of my collaborators for their insights and contributions. I am also grateful to the organizers of the Corfu workshop on Fields and Strings. Work supported by VR-grant 621-2009-4066}


\begin{thebibliography}{99}
\bibitem{Lindstrom:2004eh} 
  U.~Lindstrom,
 {\em Generalized N = (2,2) supersymmetric nonlinear sigma models,}
  Phys.\ Lett.\ B {\bf 587}, 216 (2004)
  [hep-th/0401100].


\bibitem{Lindstrom:2004iw} 
  U.~Lindstrom, R.~Minasian, A.~Tomasiello and M.~Zabzine,
{\em Generalized complex manifolds and supersymmetry,}
  Commun.\ Math.\ Phys.\  {\bf 257}, 235 (2005)
  [hep-th/0405085].
 
\bibitem{Lindstrom:2004hi} 
  U.~Lindstrom, M.~Rocek, R.~von Unge and M.~Zabzine,
{\em Generalized Kahler geometry and manifest N = (2,2) supersymmetric nonlinear sigma-models,}
  JHEP {\bf 0507}, 067 (2005)
  [hep-th/0411186].

\bibitem{Lindstrom:2005zr} 
  U.~Lindstrom, M.~Rocek, R.~von Unge and M.~Zabzine,
{\em Generalized Kahler manifolds and off-shell supersymmetry,}
  Commun.\ Math.\ Phys.\  {\bf 269}, 833 (2007)
  [hep-th/0512164].
 
  
\bibitem{Bredthauer:2005zx} 
  A.~Bredthauer, U.~Lindstrom and J.~Persson,
 {\em First-order supersymmetric sigma models and target space geometry,}
  JHEP {\bf 0601}, 144 (2006)
  [hep-th/0508228].

 
\bibitem{Bredthauer:2006hf} 
  A.~Bredthauer, U.~Lindstrom, J.~Persson and M.~Zabzine,
  {\em Generalized Kahler geometry from supersymmetric sigma models,}
  Lett.\ Math.\ Phys.\  {\bf 77}, 291 (2006)
  [hep-th/0603130].
   
 
\bibitem{Lindstrom:2007qf} 
  U.~Lindstrom, M.~Rocek, R.~von Unge and M.~Zabzine,
  {\em Linearizing Generalized Kahler Geometry,}
  JHEP {\bf 0704}, 061 (2007)
  [hep-th/0702126].
  
    
\bibitem{Lindstrom:2007xv} 
  U.~Lindstrom, M.~Rocek, R.~von Unge and M.~Zabzine,
 {\em A potential for Generalized Kahler Geometry,}
  IRMA Lect.\ Math.\ Theor.\ Phys.\  {\bf 16}, 263 (2010)
  [hep-th/0703111].
 
  
\bibitem{Lindstrom:2007vc} 
  U.~Lindstrom, M.~Rocek, I.~Ryb, R.~von Unge and M.~Zabzine,
  {\em New N = (2,2) vector multiplets,}
  JHEP {\bf 0708}, 008 (2007)
  [arXiv:0705.3201 [hep-th]].
  
  
\bibitem{Lindstrom:2007sq} 
  U.~Lindstrom, M.~Rocek, I.~Ryb, R.~von Unge and M.~Zabzine,
  {\em T-duality and Generalized Kahler Geometry,}
  JHEP {\bf 0802}, 056 (2008)
  [arXiv:0707.1696 [hep-th]].

  
\bibitem{Lindstrom:2008hx} 
  U.~Lindstrom, M.~Rocek, I.~Ryb, R.~von Unge and M.~Zabzine,
 {\em Nonabelian Generalized Gauge Multiplets,}
  JHEP {\bf 0902}, 020 (2009)
  [arXiv:0808.1535 [hep-th]].


\bibitem{Hull:2008vw} 
  C.~M.~Hull, U.~Lindstrom, M.~Rocek, R.~von Unge and M.~Zabzine,
 {\em Generalized Kahler geometry and gerbes,}
  JHEP {\bf 0910}, 062 (2009)
  [arXiv:0811.3615 [hep-th]].

  
\bibitem{Hull:2010sn} 
  C.~M.~Hull, U.~Lindstrom, M.~Rocek, R.~von Unge and M.~Zabzine,
  {\em Generalized Calabi-Yau metric and Generalized Monge-Ampere equation,}
  JHEP {\bf 1008}, 060 (2010)
  [arXiv:1005.5658 [hep-th]].


\bibitem{Hull:2012dy} 
  C.~Hull, U.~Lindstrom, M.~Ro\v cek, R.~von Unge and M.~Zabzine,
  {\em Generalized K\"ahler Geometry in (2,1) superspace,}
  arXiv:1202.5624 [hep-th].
  
\bibitem{Sevrin:1996jr} 
  A.~Sevrin and J.~Troost,
 {\em Off-shell formulation of N=2 nonlinear sigma models,}
  Nucl.\ Phys.\ B {\bf 492}, 623 (1997)
  [hep-th/9610102].
  
\bibitem{Bogaerts:1999jc} 
  J.~Bogaerts, A.~Sevrin, S.~van der Loo and S.~Van Gils,
  {\em Properties of semichiral superfields,}
  Nucl.\ Phys.\ B {\bf 562}, 277 (1999)
  [hep-th/9905141].
  
\bibitem{Buscher:1985kb}
  T.~H.~Buscher,
  {\em Quantum Corrections And Extended Supersymmetry In New Sigma Models,}
  Phys.\ Lett.\  B {\bf 159}, 127 (1985).
  
\bibitem{Gualtieri:2003dx}
  M.~Gualtieri,
  {\em Generalized complex geometry,}
  Oxford University DPhil thesis,
  [arXiv:math/0401221].
  
\bibitem{Hitchin:2004ut}
  N.~Hitchin,
  {\em Generalized Calabi-Yau manifolds,}
  Quart.\ J.\ Math.\ Oxford Ser.\  {\bf 54}, 281-308 (2003).
  [math/0209099 [math-dg]].
  
\bibitem{Gates:1984nk}
  S.~J.~Gates, Jr., C.~M.~Hull, M.~Rocek,
  {\em Twisted Multiplets and New Supersymmetric Nonlinear Sigma Models,}
  Nucl.\ Phys.\  {\bf B248}, 157 (1984).
  
  \bibitem{Courant}
  T.~Courant,  {\em Dirac manifolds}, Trans. Amer. Math. Soc., 319:631-661, (1990)

\bibitem{AbouZeid:1997cw}
  M.~Abou Zeid and C.~M.~Hull,
  {\em The gauged (2,1) heterotic sigma model,}
  Nucl.\ Phys.\  B {\bf 513}, 490 (1998)
  [arXiv:hep-th/9708047].
  
\bibitem{Gates:1984nk}
  S.~J.~Gates, Jr., C.~M.~Hull, M.~Rocek,
  {\em Twisted Multiplets and New Supersymmetric Nonlinear Sigma Models,}
  Nucl.\ Phys.\  {\bf B248}, 157 (1984).



 
\end{thebibliography}
\end{document}